# Micro- and Nanostructured Diamond in Electrochemistry: Fabrication and Application

Fang Gao, Christoph E. Nebel

Fraunhofer Institute for Applied Solid State Physics, Freiburg, Germany

## Introduction

Synthetic diamond has been studied extensively during the last 30 years due to its outstanding chemical/physical properties such as wide bandgap, high thermal conductivity, and extreme hardness. Compared to bulk diamond, diamond materials with micro- and nanostructured surfaces provide a facile way to fine tune diamond properties such as field emission, hydrophilicity, and specific surface. In this chapter, the fabrication and application of micro- and nanostructured diamond will be discussed.

The fabrication method of diamond nanostructures can be divided into two categories: top-down etching and bottom-growth. The early work on 3D micro-structured diamond dates back to mid-1990s, using chemical vapor infiltration (CVI) techniques. In this technology, carbon or carbide fibers were typically used as the growth template. Almost in parallel, reactive ion etching (RIE) was applied to achieve diamond surface nanostructuring. After that the diamond surface nanostructures, typically vertically aligned diamond nanowires (or nanorods) has been mainly fabricated using top-down plasma etching techniques. In recent year, the templated diamond growth has gained increasing attention due to the wide choice of template, mask-free production, and unlimited surface enlargement. In this chapter, the development and main techniques used in these two approaches will be elaborated. Nevertheless, other less common methods, such as catalytic etching by metal particles, steam activation and selective materials removal will also be discussed.

As indicated by the title, this chapter will mainly deal with the application of micro- and nanostructured diamond in electrochemistry. In these applications, the advantage of nanostructured diamond can be divided into three aspects: 1) providing enlarged surface area for charge storage and catalyst deposition; 2) tip-enhanced electrochemical reactions used in sensing applications; 3) diamond membranes with micro- or nanopores can be applied in the electrochemistry separation and purification applications. Examples and explanations on these applications will be given in this chapter.

## Fabrication method of diamond nanostructures

### Reactive Ion Etching

Reactive ion etching is a dry etching technique using reactive plasma to remove material from a surface. Schematic illustrations of such techniques are shown in **figure 1**. The plasma is ignited either capacitively (**figure 1a**) or inductively (**figure 1b**). In both cases, a radio frequency (RF) voltage is added between the anode (the top and the wall of the reaction chamber) and cathode (the sample platter) to accelerate the electrons up and down; in the meantime, the massive ions are relatively unaffected. When electrons hit the anode, they are fed out to the ground. However, because the sample platter is (direct current) DC insulated, a negative potential will build up when electrons touch the cathode. In the capacitively couple plasma (CCP) etching, the energy of plasma is coupled with the negative bias of the sample, i.e. when a high-density plasma is built, a high bias is unavoidable. As a result, the freedom to adjust the etching parameters is very limited. In the latter case, however, the plasma energy is decoupled from the sample bias. Therefore, high-density plasma is allowed without the danger to overcharge the sample.

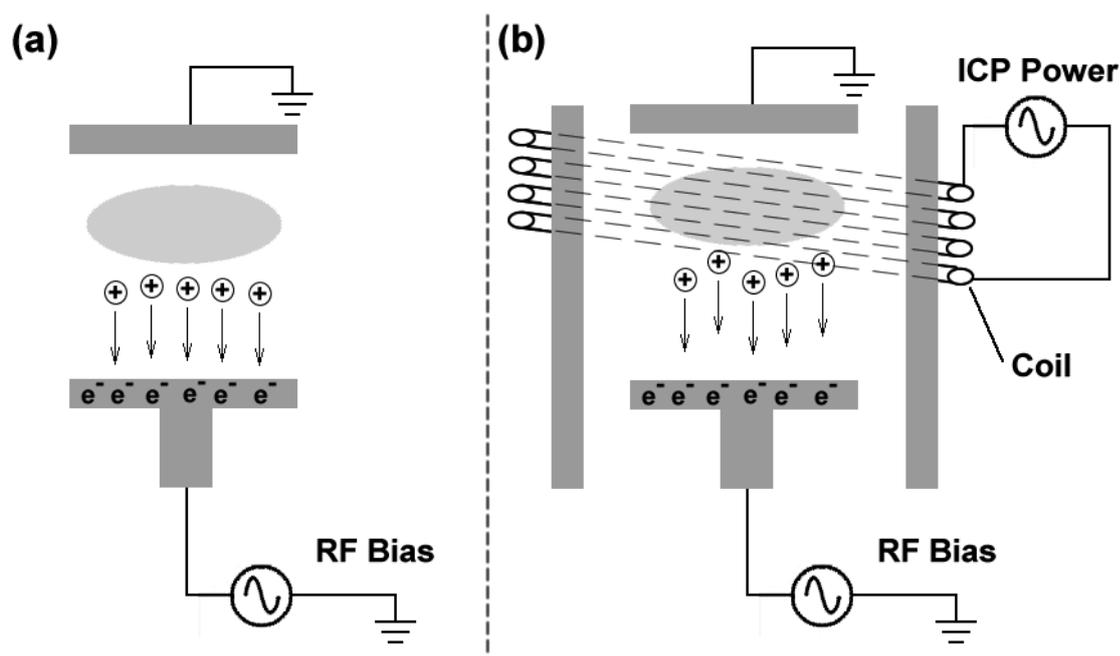

**Figure 1.** Schematic illustration showing different RIE mechanisms: (a) capacitively coupled plasma etching; (b) inductively coupled plasma (ICP) etching.

The RIE techniques on diamond were developed in late 1990s. in 1997 Shiomi reported the diamond surface etch via $CF_4/O_2$ plasma [1]. Using sputtered Al with a thickness of 0.4 µm as the etching shadow mask, diamond surface microstructures were formed. In this research, it was found out that the ratio of $CF_4$ and $O_2$ in the etching atmosphere plays an important role in the etching selectivity between diamond and Al. however, an abnormal phenomenon was also reported: when pure $O_2$ plasma was used to etch polycrystalline diamond (PCD), a high density of diamond nanowires will appear. The explanation at that time was local surface phase transition induced by ion bombardment introduced self-masking effect to the etching process. These mask-free or self-masking phenomena were repeatedly used in the diamond nanowire formation processes [2, 3]. However, more recent research has shown that these effects are likely due to the oxide residues in the etching chamber acting as the etching mask, because high-resolution transmission electron microscopy (HRTEM) has revealed the existence of a thin layer of amorphous oxide deposited all over these nanowires [4].

Metal nanoparticles deposited on diamond surface are often used as etching masks. In 2008, Zou et al reported the formation of gold nanoparticle on diamond surface via a thermal dewetting method [5]. The diamond sample coated by 5 nm Au layer was heated up to 850 °C in Ar/$H_2$ plasma. The Au nanoparticles were formed in situ via the self-organization of the molten thin layer, and they were used as the shadow mask in the later RIE etching. In this method, diamond nanopillars of a high density of ~ $10^9$ cm$^{-2}$ were formed with diameters of ~ 30 nm and heights of ~ 400 nm. A higher density and aspect ratio was later achieved by using more resistive etching masks and more anisotropic etching techniques. In 2010, Smirnov et al, reported the high aspect ratio (~ 30) and high density diamond nanowires fabricated using Ni nanoparticles as etching mask and ICP etching techniques [6]. The method is shown in **figure 2 (a)**: a thin Ni film of 1 nm was evaporated on the PCD surface, and the layer was later thermally melted to form a dense layer of nickel particles (~ $10^{10}$ cm$^{-2}$, **figure 2b**). The diamond wires were formed by using these particles as shadow mask in an ICP etching process (**figure 2c**). Finally the mask residues were removed by wet-chemical cleaning. Dimensions of nanowires were: height (1200±200) nm, width (35±5) nm, density ~$10^{10}$ cm$^{-2}$.

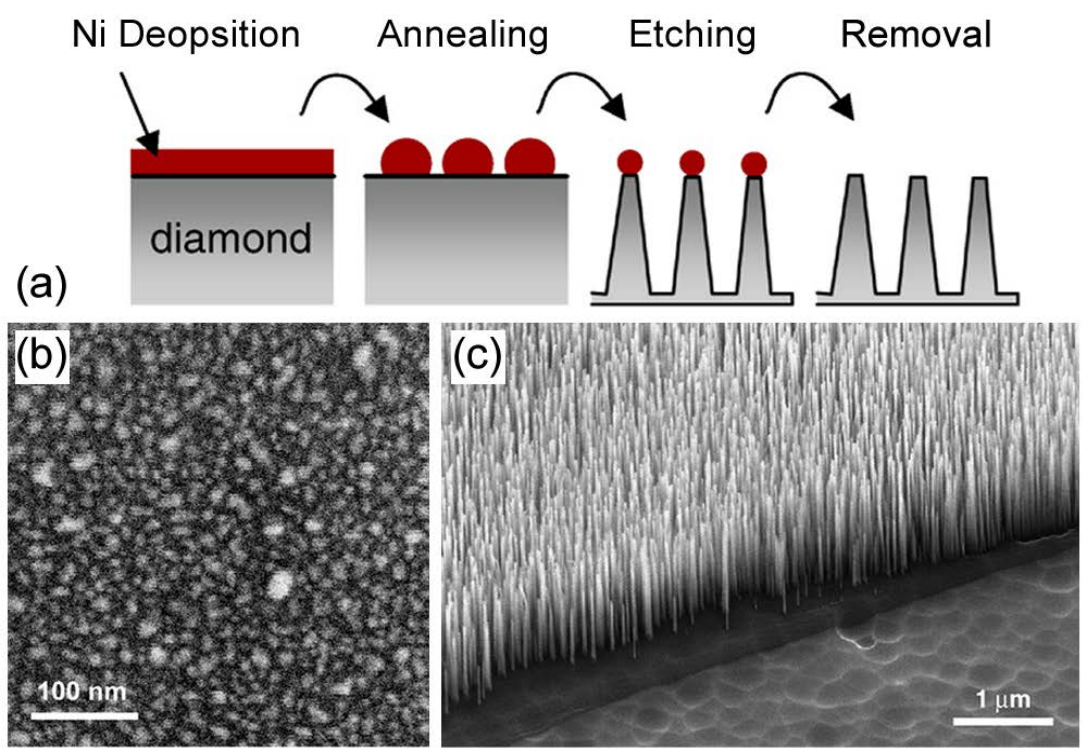

**Figure 2.** (a) Scheme showing the fabrication method of diamond nanowires; (b) an SEM image showing nickel nanoparticles from the dewetting of a 1 nm thick nickel film on diamond; (c) an SEM image showing the diamond nanowire after etching and nickel removal.

Electrochemical deposition provides another method to controllably deposition metal nanoparticle as mask for etching processes. Gao et al has shown that the density and size of the Pt nanoparticles electrochemically deposited on diamond electrode surface can be fine-tuned via surface pretreatment and deposition time [7]. The density is tunable in the range of 3.3±0.5 – 12.7±1.5×$10^9$ cm$^{-2}$, and the size can be adjusted between 18±5 and 46±9 nm. These particles can be used as etching masks for diamond column fabrication [8].

Alternatively, mask can also be formed using self-organized colloid particles. Periodically organized 2D structures of spherical particles can be obtained using capillary force and water evaporation [9, 10]. Okuyama et al reported the fabrication of well-ordered diamond micropillars using self-organized $SiO_2$ spheres [11]. In this research, $SiO_2$ particles of 1 µm diameter was coated on the flat nucleation side of the PCD sample, forming a self-organized monolayer of a 2D hexagonal close-packing of $SiO_2$ spheres. By consequential $O_2$ plasma etching, this structure is transferred onto the underlying diamond substrate. Similarly, Yang et al used self-organized diamond nanoparticles as etching masks [12]. In this case, surface dipole interactions between the diamond seeds and the diamond substrate result in the attachment of seeds onto the substrate [13]. The diamond seeds with a typical size of 8 – 10 nm were dispersed in water and treated with ultrasonication. The diamond to be etched was then immersed in the suspension. The density of the particle attached to the diamond surface can be adjusted by the immersing time and the concentration of the suspension. After nanoparticle attachment, the diamond surface was etched in oxygen-rich plasma. In this process, diamond nanowires of 10 nm length and an average separation of 11 nm were obtained with an etching time of 10 s.

Besides nanowires, nanoporous materials can also be fabricated using top-down etching method. RIE with shadow mask is a pattern transferring technique in principal. Therefore, more sophisticated nanoporous structures can be obtained if a nanoporous mask is used. For example, porous anodic aluminum oxide (AAO) has been studied for more than 50 years [14]. The pore size and distribution can be well-controlled by solution composition, applied potential and temperature [15]. It has been widely used as shadow masks for metal deposition[16] and pattern transfer on Si [17]. In 2000, Masuda et al developed a fabrication method of diamond honeycomb electrode using porous AAO as the etching mask [18]. In the later work, the same group also realized and characterized porous diamond electrodes with a wide range of pore sizes and depths using this method [19, 20]. By etching through the diamond layer, diamond membrane was also fabricated [21].

### Templated-Growth
On the contrary to top-down etching method diamond surface can also be fabricated via templated-growth. In the 1990s in order to enhance the nucleation density, people used porous silicon as the growth substrate [22-25]. After that other porous materials, such as porous titanium[26], Zeolite [27] and porous carbide [28] were also used in for these reason. Normally, planar films were obtained. However, porous diamond films are sometimes obtained [29-31]. In these method, the porous substrate are used, no matter intentionally or not, as the growth template. The diamond film grown on top of the porous substrate inherited the porous structures. A more typical templated-growth method came in the mid-1990s. Chemical vapor infiltration (CVI) was used in this method [32, 33]. In CVI, microwave plasma or a hot-filament will generate reactive species above the porous substrate. Chemical vapor deposition (CVD) in porous structures is enhanced by the manipulated gas flow which send the carbon and hydrogen radicals through the porous sample. Although there are successful reports on diamond coating up to millimeter-depth into the porous template [34], there is no evidence showing that this technique is able to coat nanoporous materials. A more "modern" growth method came in 1999. Demkowicz et al report the plasma CVD growth of diamond on a SiC-whisker compact [35]. In this research, a diamond seeding procedure was performed by adjusting the pH to 12.7.

Diamond seeds with diameters of ~300 nm were attached to SiC whiskers for further overgrowth. With a coating thickness of ~300 nm, the diamond coating was formed up to a depth of ~50 μm. In 2001, Baranauskas et al reported the templated diamond on natural pyrolized fibers [36]. In their research they compared the result of overgrowth with and without nanodiamond seeding process, and showed the necessity of the seeds for a complete coating.

Later developments of the 3D diamond coating follow the similar approach; the focus is mainly on tests on a large variety of materials for the growth template. Si-based materials due to their high melting point, chemical inertness, stable interface with carbon, as well as the suitable thermal expansions are widely applied as the growth template. In 2009, Luo et al reported the growth of diamond on silicon nanowires using hot-filament CVD method [37]. The wires are vertically aligned and the coating covers the entire wires (~5 μm). In the same year, Kondo et al, reported the diamond fiber growth on quartz fiber filter [38]. Compared to the results of Luo et al, a more complex template with interweaved $SiO_2$ fibers was used. The growth was carried out at a very low temperature of 500 °C. The filter paper used has a thickness of 450 μm. However, the diamond growth only penetrated the top 15 – 20 μm of the template. Also, the coating shows a non-homogenous nature: the coating thickness at the surface can be one order of magnitude higher than deep in the template.

Nanomaterials based on $sp^2$ carbon are also candidates for the templated growth. Already in 2005, Terranova et al reported the nanodiamond coating on carbon nanotube (CNT) [39]. The meaning of this work is limited by the fact that they were using a specially designed CVD system with nanocarbon as the carbon source. After that there has been no further report on diamond/CNT core-shell structures for more than five years. The difficulty in coating CNT with diamond lies in two aspects. The first is the difficulty for seeding. Hydrocarbons are known to be mostly hydrophobic. Therefore the water-based diamond colloid cannot properly penetrate. The second problem is more fundamental. The typical diamond growth condition requires the hydrogen plasma etching of the co-deposited graphitic carbon. Therefore, the CNT which acts as the growth template can be etched during the growth as well [40]. In 2012, Zou et al reported the diamond/CNT teepee-like composite using optimized seeding and growth techniques [41]. In this study, the nanodiamond seeding was carried out using electrospray method. The methanol suspension of 5 nm nanodiamond was electrosprayed under 35 kV bias on to the grounded CNT substrate. After drying, the CNT array showed a teepee-like morphology. The diamond was deposited using 1% $CH_4$ in $H_2$ and ~1200 K.

Besides wire/fiber templates spherical compact from silica or opal lattice is also often used as the growth template. In 2012, Kurdyukov et al reported the fabrication of porous diamond membranes using templated growth on synthetic opal film consists of $SiO_2$ beads (diameter: 520±30 nm) [42]. The template lattice was formed via self-organization driven by the capillary force. To further enhance the stability of the lattice, high temperature (1000 – 1050 °C) annealing was applied. In this research, up to 15 layers of silicon dioxide spheres were coated. Similar work has been reported by Kato et al in the same year on boron-doped diamond foam for electrochemistry applications [43]. The template was fabricated by simply drop-casting $SiO_2$ spheres on boron-doped diamond substrate. The boron-doped diamond foam showed a strong Fano resonance in the Raman spectrum, which indicated a metallic conductivity [44, 45].

Also, a redox peak separation of ~ 59 mV for Hexaammineruthenium chloride is recorded, which shows that this material is suitable for electrochemistry.

Currently, the nonuniformity of the diamond coating in a 3D template is one of the most prominent problems for this growth technology. In order to obtain larger surface area in one growth, the diamond coating needs to penetrate as deep as possible. However, there is a contradiction in nature. If the radical chemistry in diamond growth is considered, one would find the following formula [46]:

$$C_DH + H^· \leftrightarrow C_D^· + H_2 \qquad (1)$$

$$C_D^· + CH_3^· \leftrightarrow C_DCH_3 \qquad (2)$$

Formula (1) is the activation process of a surface C-H site, and the formula (2) is the addition of a $CH_3$ group on the diamond surface. For the normal growth conditions, atomic hydrogen and methyl radicals need to be abundant. On a planar sample surface, this is normally the case. In 3D templates, however, this requirement is hardly satisfied. The reason is that when the radicals collide into the template, they lose the kinetic energy which is necessary to trigger the above mentioned reactions. For this reason, the diamond growth will decrease deeper into the template (figure 3).

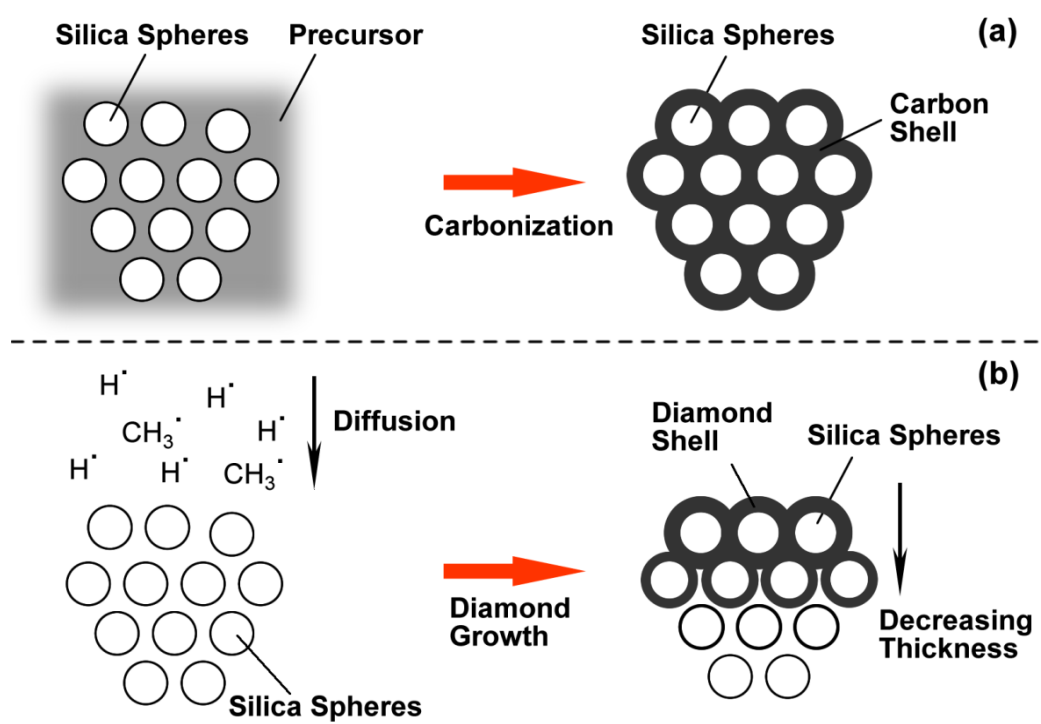

**Figure 3.** Schematic illustrations showing the difference between templated-growth of porous $sp^2$ carbon (a) and the templated-growth of diamond (b). Reprinted with copyright permission from ACS Applied Materials Interfaces, p. 28244-28254 (2016) DOI: 10.1021/acsami.5b07027

Concerning this problem, there is up to now no theoretical optimization about the growth condition regarding plasma power, gas pressure, methane concentration as well as the sample temperature. However, most recent works concentrated on lower temperature (< 600 °C) lower pressure (< 30 mbar) growth conditions. In these conditions, the diamond growth is slow, and the mean-free-path for radicals is longer. Moreover, if the diamond growth is slow, the upper pores of the template will be closed slower, so that the growth on the lower parts will be less hindered. Also, by using low temperature more choices of template materials are available. In 2015, Ruffinatto et al reported the coating of glass fiber filter paper up to a thickness of ~250 µm [48]. This deep thickness is possibly due to the optimized growth condition and the large (micrometer-sized) pores in the template. In the same year, Hébert et al reported the coating of nanoporous conductive polymers using nanocrystalline with an ultralow-temperature growth technique (<450 °C) [49]. In this study the authors emphasized the importance of high seeding density achieved by infiltration method. They claim that the seeds layer can prevent the etching effect of the hydrogen plasma on the polymer. The electrochemistry measurements showed that the background redox activities of the underlying porous polypyrrole layer were completely quenched after the diamond coating, showing that the coating is of pinhole-free quality.

Beside the optimization of the diamond growth parameters, another way to overcome the penetration problem in the templated growth is the layer-by-layer growth method [47]. Gao et al reported an improve fabrication method of the diamond foam fabrication reported in ref. [43]. Rather than depositing the template and diamond in one deposition, they deposit the composite in a layer-by-layer manner (**figure 4**). In this method, the diffusion limitation shown in **figure 3** is relieved, and the thickness of the diamond has no theoretical upper limit.

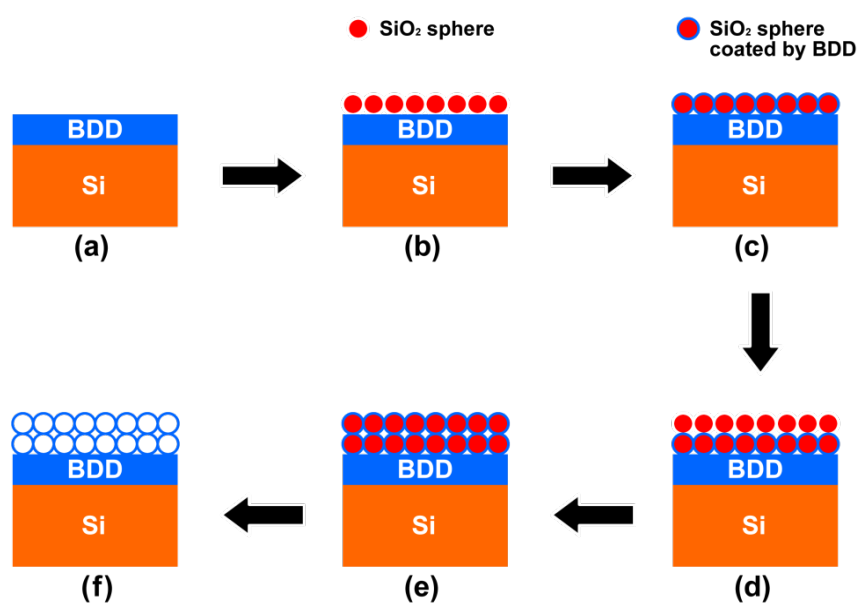

**Figure 4.** Schematic illustration showing the layer-by-layer growth technique for diamond foam electrodes showing: (a) boron-doped diamond (BDD) substrate growth on Si; (b) spin-coating of $SiO_2$ spheres on the substrate; (c) CVD diamond coating on $SiO_2$ templates; (d) spin-coating of the second $SiO_2$ layer, (e) the second CVD diamond coating on $SiO_2$ templates; (f) removal of $SiO_2$ templates. Reprinted with copyright permission from ACS Applied Materials Interfaces, p. 28244-28254 (2016) DOI: 10.1021/acsami.5b07027

Finally, there is a kind of less typical templated growth technique which can be called masked-growth. It resembles the templated growth in the way that a porous template is needed; however, the diamond will not grow on the template but on the unmasked area. Such methods are very similar to the metal deposition with shadow masks which has been often applied to obtain highly ordered metal dot patterns using ordered particles or porous AAO [16, 50]. In 2001, Matsuda et al fabricated vertically aligned diamond nanocylinders using this method [51]. In this research, porous AAO was used as the mask for diamond cylinder growth. The bottom of the through-hole AAO film was seeded with nanodiamond (size: 50 nm). Afterwards, the AAO film was flipped over and the CVD deposition was carried out from the front side. The result showed that well-aligned polycrystalline diamond nanocylinders with a density of $4.6 \times 10^8$ cm$^{-2}$ were generated. The morphologies including the hexagonal close-packing and the average diameter were both inherited from the AAO mask, showing the effect of masked growth. By properly shaping the AAO masks, diamond cylinders with triangular and square cross sections can also be synthesized [52].

## Surface Anisotropic Etching by Metal Catalyst

Diamond can react with $H_2$ at high temperatures with the help of metal catalyst particles. The etching mechanism was reported as early as 1993 by Ralchenko et al [53]. They reported on the diamond patterning using Iron group elements (Fe, Co and Ni). The etching mechanism was explained in three steps: 1) carbon dissolution in metal, 2) diffusional transport to the metal-gas interface and 3) carbon desorption in the form of methane. They also showed that Fe has the strongest etching effect. This phenomenon was later discovered in nanoscale by Konishi et al in 2006 [54]. They found that the Co nanoparticles generated by in-situ reduction from $Co(NO_3)_2$ in $H_2$ atmosphere could catalytically etch diamond surface in the same atmosphere and leave nanosized etch-pits on diamond surface. They also found that the geometry of the etch-pit is facet-dependent: on (111) facets the etch-pits were triangular or hexagonal (**figure 5a**); on (100) facets, the etch-pits were rectangular (**figure 5b**); on (110) facets, channels along the {111} direction will be formed (**figure 5c**). This nanopit formation phenomenon was also reported later on other metal particles, including Ni [55], Fe [56, 57] and Pt [58]. Au nanoparticles are reported to be unable to etch diamond surface under heating and $H_2$ atmosphere [59].

In the fabrication of nanoporous structures using catalyzed etching, the density of the etch-pits is dependent on the density of metal nanoparticles. Therefore, particle density is an important parameter to control. Mehedi et al have shown that if the particles are formed by thermal dewetting of a thin Ni metal film, the density of particles is linearly decreasing with the increasing thickness for films thinner than 5 nm [60]. However, the difference is within one order of magnitude ($10^{10}$ cm$^{-2}$). For thickness more than 5 nm, the density drops heavily. Also, they found out that the etching stops at a depth of 400 – 500 nm into the diamond surface, due to the lack of hydrogen inside the deep pores [59].

A detailed study on the etching mechanism is published recently in ref. [60]. In this study, diamond substrates were etched by Ni nanoparticles during a $H_2$ annealing process. The surface chemical composition was monitored via X-ray photoelectron spectroscopy (XPS) and electron energy loss spectroscopy (EELS) throughout the annealing process. No formation of nickel carbide was detected both before and after Ni removal. Moreover, the gas composition in the reaction chamber is also analyzed. There was a clear increase in the amount of methane in the gas mixture, and this increase is in accordance with the carbon loss calculated from the average depth and diameter of the nanopores. Therefore, the three-step etching mechanism is confirmed.

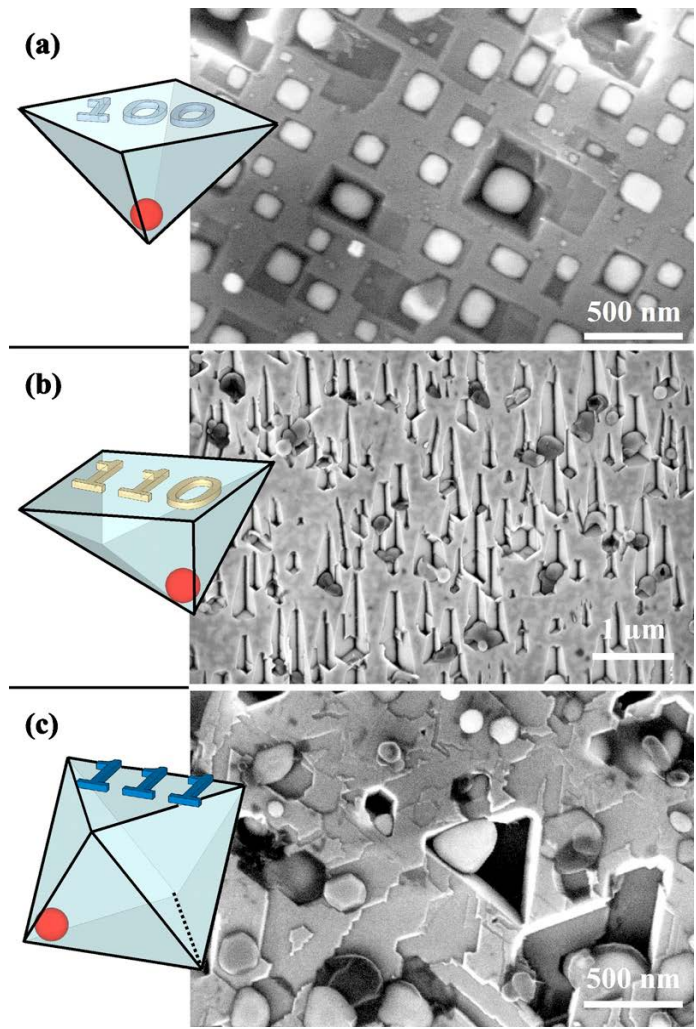

**Figure 5.** SEM images of observed pitting and channeling of (a): (100), (b): (110) and (c): (111) directed and etched single-crystal diamonds. The inset is a modeled octahedron which reflects symmetries of {111} oriented planes with indicated planes. Red circles indicate Ni. Reprinted with copyright permission from Appl. Phys. Lett. 97, 073117 (2010); doi: 10.1063/1.3480602

### High-Temperature Surface Etching

Even without metal catalyst diamond can react with $H_2$, water vapor and $CO_2$ at high temperatures. This phenomenon is discovered when researchers tries to understand the surface etch-pits (trigons) on nature diamonds [61]. There have been positive and negative etch-pits. Their formation and their relation with a variety of crystal defect have been reported [62]. However, because these etch-pits are shallow and micrometers in size, their effect on the local properties and the total surface area is small. Therefore, we will not go into the details of these effects in this chapter.

A practical catalyst-free etching or surface roughening method of diamond came in 2011. Ohashi et al reported the diamond surface nanotexturing via the steam-activation method which is normally applied to activated carbon [63]. Temperatures between 600 and 900 °C were applied to the sample. Water vapor was used as etchant. The etching effect started at 700 °C: some triangular shaped etch-pits appeared on (111) facets. At higher temperatures rigorous corrosion of the BDD surface was observed. High density of columnar structures was obtained on the diamond surface. From the analysis of the outlet gas, CO and $H_2$ were confirmed to be the product. Therefore the etching mechanism is presumably:

$$C + H_2O \leftrightarrow CO + H_2 \qquad (3)$$

An interesting consequence of the etching is the enhancement of the peak ratio between the diamond peak and the graphitic peak, showing a decreasing amount of $sp^2$ contentafter the process. Together with the observed fact that (111) facet was more prone to etching, a hypothesis about the etching mechanism was deduced: the activation process first state with the $sp^2$-rich grain boundaries, and then the (111) facets which are close to grain boundaries are also selectively etched following the reaction (3). Some recent results published by the same group shows that boron-concentration has also an effect on the surface enlargement of diamond [64]. It is believed that in highly boron-doped samples, the proportion of (111) is larger than in lower boron-doped samples. Therefore, the etching has a stronger effect of the surface.

Other gases such as $CO_2$ and $O_2$ have also shown similar effects. Zhang et al has reported that the activation process with $CO_2$ atmosphere at 800 and 900 °C has a preferential etching on the (100) facets [65], via:

$$C + CO_2 \leftrightarrow 2CO \qquad (4)$$

Because $O_2$ is a much stronger oxidant than $H_2O$ and $CO_2$, direct annealing in $O_2$ containing atmosphere will have an uncontrollable etching effect on the sample in a relative short period. Therefore, special procedures are needed if $O_2$ is used for diamond etching. Kondo et al shows that if this process is split into two phases, i.e. a high temperature (1000 °C) graphitization process in Ar and a mild (425 °C) oxidation in air, controlled and selective etching will happen on both (111) and (100) facets on the diamond surface [66].

### Selective Material Removal

It has been shown for a long time that diamond is almost nondestructible wet- or electrochemically [67]. Diamond films can be used in acidic fluoride [68], alkaline [69]. Electrochemically, diamond has been used in the mixture of 1.0 M $HNO_3$ and 2.0 M NaCl at a current density of 0.5 A cm$^{-2}$ for up to 12 h with no evidence of damage [70]. In $H_2SO_4$ solutions diamond electrodes have shown stability in a current range of 1 – 10 A cm$^{-2}$ [71]. These information indicates the possibility to grow diamond together with some other materials and selectively remove the co-deposited materials after deposition. In this way, a porous diamond backbone will be fabricated.

The most commonly deposited non-diamond material in diamond growth is graphite. In the early research on diamond growth, it was known that graphitic carbon is co-deposited with diamond in CVD process [46, 72]. Normally, this graphitic growth needs to be minimized to enhance diamond quality. However, if a diamond sample is intentionally deposited with a high $sp^2$-content, and the non-diamond carbon is removed afterwards via selective etching, a way towards porous diamond can be found. This is shown by Kriele et al in 2011 [73]. Using up to 20% methane in the gas mixture for diamond growth, highly graphitic diamond is deposited. Raman spectroscopy shows a very high G and D band showing the poor quality. By partially removing the $sp^2$ content in air at ~550 °C nanopores can be generated on the diamond film. In this work, 150 nm thick diamond films were used. As a result, a nanoporous membrane was generated. Similar work was reported by Feng et al [74]. However, they reported the selective removal from a thicker diamond film. The results showed that after etching away the non-diamond carbon, a fibrous diamond skeleton was formed. TEM shows that these fibers consisted of fine diamond grains. While the outer shape of the micrometer-size grains was kept, the diamond surface becomes highly porous.

The graphite removal can also happen in situ if etching gases are introduced in the growth atmosphere. In this method, diamond growth took place at the edge of no-growth region of C–H–O ternary diagram [75]. Diamond porous nanowire structures were generated directly from the growth. The authors used very different gas mixtures from a "typical" diamond growth: very high methane concentration together with a high percentage of $CO_2$ ($CH_4$: $CO_2$: $H_2$=0.2: 0.8: 1). In this combination, the authors believed that the $sp^2$ carbon deposited by the high methane concentration was etched in situ by the excessive $CO_2$. This process results in very porous surface structures.

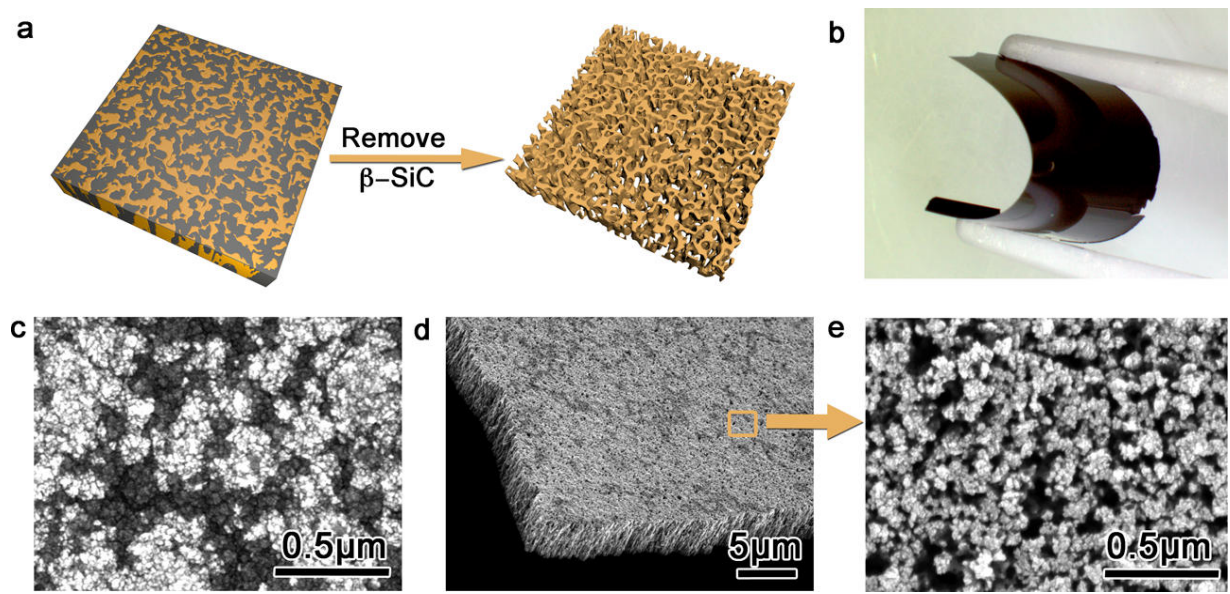

**Figure 6.** (a) Illustration of fabricating a diamond network from the composite film: the gray phase (β-SiC) is removed, leaving a yellow porous phase (diamond). (b) An optical photo of a flexible freestanding diamond network film. (c) SEM surface images of a nanocrystalline diamond/β-SiC composite film deposited with TMS/CH4 ratio of 1.5%. (d) Diamond network fabricated by etching the β-SiC phase from composite film shown in image c. (e) High-magnification SEM images of the surface of the film shown in image d. Reprinted with copyright permission from ACS Applied Materials and Interfaces, 5384-5390 (2015) DOI: 10.1021/am508851r

Recently, a new approach was reported by scientist working in the SiC field. Rather than depositing diamond and remove the by-product after the growth, they grow SiC and keep the by-product, which is diamond. Zhuang et al reported the fabrication of free-standing diamond network by the selective removal of *β*-SiC in a SiC-diamond composite via wet-chemical etching (**figure 6**) [76]. They varied the gas phase ratio between $CH_4$ and tetramethylsilane (TMS) to obtain different porosities in the network. When TMS/$CH_4$ ratio varies between 0.8 and 2.2% the porosity shifted from ~15 to ~70% almost linearly. This method shows a powerful tool in the fabrication of diamond porous membranes. From another view point, it is a kind of templated growth where the template is grown in situ. Therefore, the problem mentioned in **figure 3** is solved.

## $sp^2$-Carbon Assisted Growth of Diamond Nanostructures

In some occasions, diamond nanostructures can be co-deposited with $sp^2$ carbon in some "unusual" deposition conditions. For all these conditions, the $sp^2$ carbon grows simultaneously with diamond, and the non-diamond carbon acts as the "template" for the diamond growth. In 2004, Sun et al reported the growth of diamond nanorods during the $H_2$ plasma post-treatment of CNT [77]. This discovery was based on their earlier discovery of a high density nucleation on CNT during short time (<10 h) $H_2$ plasma treatment [78]. They found that if the treatment time was elongated to > 20 h, diamond nanorods with diameters of 4 – 8 nm and length up to 200 nm were synthesized. Based on the observation that these nanorods are covered by a thin layer of amorphous carbon sheath, they deduced a possible growth mechanism (**figure 7**): the amorphous $sp^2$ carbon clusters were generated first in the plasma. By continuous insertion of hydrogen, diamond nucleates were generated inside the clasters, which is followed by further crystal growth. During the diamond growth, the competing deposition of amorphous carbon will continue and wrap the outer surface of the as-grown diamond structures. In this mechanism, the amorphous carbon sheath plays a decisive role in the 1D growth of the diamond wire by confining the lateral growth.

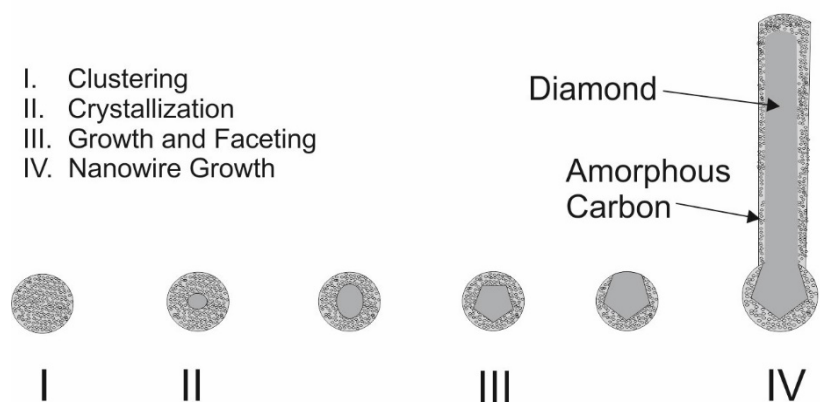

**Figure 7.** The proposed model for the formation of nanodiamonds, and the growth of diamond nanorods under hydrogen plasma irradiation of MWCNTs at high temperatures. Amorphous carbon clusters are formed in step I. The crystallization of diamond begins in the core of the carbon clusters (step II), followed by the diamond growth and faceting stage (step III). After the diamond nanocrystallites are faceted, diamond nano-rods begin to grow at the nanorod tips (step IV).

A similar but more controllable growth method came later in 2010. Hsu et al reported the diamond growth in atmospheric pressure CVD condition which is commonly used for CNT growth [79]. Similar to the previous work, they discovered the growth of diamond wires inside a CNT sheath. However, they wires are 60 – 90 nm in diameter and up to tens of micrometers in length. In this research, the importance of atomic hydrogen is also stressed. It is reported that the 12 h cooling process (1.2 °C min$^{-1}$) was indispensable to for the diamond wire synthesis.

In a recent report by Zhang et al the diamond has been "assembled" in side double wall CNT using diamantane dicarboxylic acid (DDA) as the building block [80]. The assembly process was realized by pulling individual DDA molecules inside the CNTs by a force resembling the capillary force, which is monitored and proved by high resolution transmission electron microscopy (HRTEM). The carboxylic groups in DDA were later removed by $H_2$ annealing at 600 °C for 12 h. In this way, the diamantane part of DDA is connected forming a diamond nanowire with a diameter of 0.78 nm.

Besides the CNT-related growth, a $N_2$ containing deposition gas mixture will also generated wire-shaped morphology of the resulting diamond film. In 2007, Arenal et al carried out a thorough research on ultrananocrystalline diamond (UNCD) growth under $N_2$ containing plasma [81]. They observed the formation of diamond nanowires with high density and uniform distribution at temperatures > 800 °C and with $N_2$ proportions >10% in the reaction gas mixture. HRTEM results showed that for the samples grown at 10% $N_2$, 800 °C, the nanowires formed are 80–100 nm long with a core-shell structure. The diamond core is composed of 5 nm wide, 6–10 nm long nanocrystalline segments. This core is enveloped by $sp^2$ carbon, which is similar to the CNT-related cases. Almost the same results were reported in the same year by Vlasov et al during the N-doped UNCD growth [82]. Also, similar to ref [77], these wires were elongated in the (110) direction, showing similar growth mechanism. In fact, the (110) preferential growth in $C_2$-dimer-dominated UNCD growth is confirmed theoretically [83]. The influence of $N_2$ concentration is likely to be the result of CN dimers which can preferentially attach to certain diamond facets [84].

### High Pressure High Temperature (HPHT) Methods

Typically, the above mentioned techniques for porous diamond fabrication are based on CVD growth, or based on bulk diamond fabricated by CVD. Meanwhile, HPHT method has also been used to fabricate porous diamond. There are two approaches reported: (1) Starting from porous $sp^2$ carbon materials and using HPHT treatment to turn $sp^2$ into $sp^3$ carbon; in 2011, Zhang et al fabricated monolithic transparent porous diamond crystals from mesoporous carbon CMK-8 via HPHT treatment (21 GPa, 1600 °C) [85]. The aim of this research is to lower the $sp^2$-$sp^3$ conversion temperature by using mesoporous carbon. The diamond product inherited the porous structure, although specific surface of the porous diamond obtained was 33 m$^2$ g$^{-1}$, much smaller than the starting porous carbon (1250 m$^2$ g$^{-1}$). (2) Using HPHT treatment to sinter diamond powder into porous bulk ceramic. Due to the high melting point of diamond, the sintering of diamond particle takes place also at high temperatures. In 2009, Zang et al reported the fabrication of bulk BDD electrode using BDD particles [86]. Sintering was carried out at 1450 °C, 6 GPa, with 15 wt% Fe–Co–B alloy powders as sintering catalyst. The BDD ceramic has 1-10 μm pores associated with grain boundaries; the porosity is measured to be 14%.

# Application of diamond nanostructures in Electrochemistry

## Biosensors based on nanostructured diamond

The Biosensors is an important application for diamond-based nanomaterials. The appealing properties of diamond for biosensing include bio-compatibility [87-89], facile surface termination [90, 91] and functionalization [92-95], and easily cleaned and restored surface [96]. This application starts in 2008 with the work from Yang et al using boron-doped diamond nanowires as a DNA sensing platform (figure 8) [97]. Slightly earlier, they discovered that the electrochemical surface grafting of diamond nanowires with nitrophenyl linkers took place preferentially at the top of the wires [12]. Therefore, they continued the research by attaching single strand DNA (SS-DNA) on the top of the aminophenyl linker. The redox responses from ferro/ferricyanide are highly sensitive to the surface condition of the diamond nanowires. The peak current shrinks gradually as the phenyl linker (**figure 8c**), cross-linker (**figure 8d**), SS-DNA (**figure 8e**), and the complementary DNA (which forms double strand DNA with the immobilized SS-DNA **figure 8f**) were attached to the surface. With this technique, the detection limit for DNA is lowered to ~2 pM with good reproducibility and anti-interference properties against mismatching SS-DNA [98, 99].

Besides DNA sensing, diamond nanostrutures are also used in the sensing of a wide varieties of bio-organic chemicals. In 2009, Wei et al reported the diamond grass electrode had an improved sensitivity towards the detection of uric acid and dopamine [2]. The electroxidation current was recorded from the electrode. Compared to a flat BDD electrode, the current as well as the reaction kinetics are both enhanced after surface modification. It is believed that the nanowire formation enhance the number of reactive site at the electrode surface (e.g. the tip of the wires), and thus a larger signal current.

In some cased, surface nanostructures do not only multiply the signal according to the surface-enlargement but also enable the electrode to detect substances which are otherwise not detectable. For glucose sensing, the flat BDD electrode is report to be nonreactive, the electroxidation current appear only after nanostructuring [37, 100]. Luo et al reported the amperometric glucose sensor using diamond-coated silicon nanowires [37]; a sensitivity of 8.1 μA mM$^{-1}$ cm$^{-2}$ with a limit of detection of 0.2±0.01 μM was achieved. In a later research, diamond nanowires electrode was also used to detect tryptophan using differential pulse voltammetry [101]. The detection limit was $5 \times 10^{-7}$ M was obtained on BDD nanowires, as compared to $1 \times 10^{-5}$ M recorded on planar BDD electrodes [102].

It is worth noticing that voltammetric methods have seldom used in these detection. The reason is probably because of the heavily increased background current after the surface nanostructuring (we will come to this point again in the next section). In voltammetry techniques, the signal is limited by diffusion. Therefore, a further increase of the surface only increases the capacitive background proportional to the surface-enlargement but not the signal. Therefore, even if voltammetric method has been used, the background current need to be suppressed with pulse techniques [101]. This has been pointed out in the very early days of porous diamond electrode research [20] and repeatedly confirmed in later researches [6, 43, 49, 103].

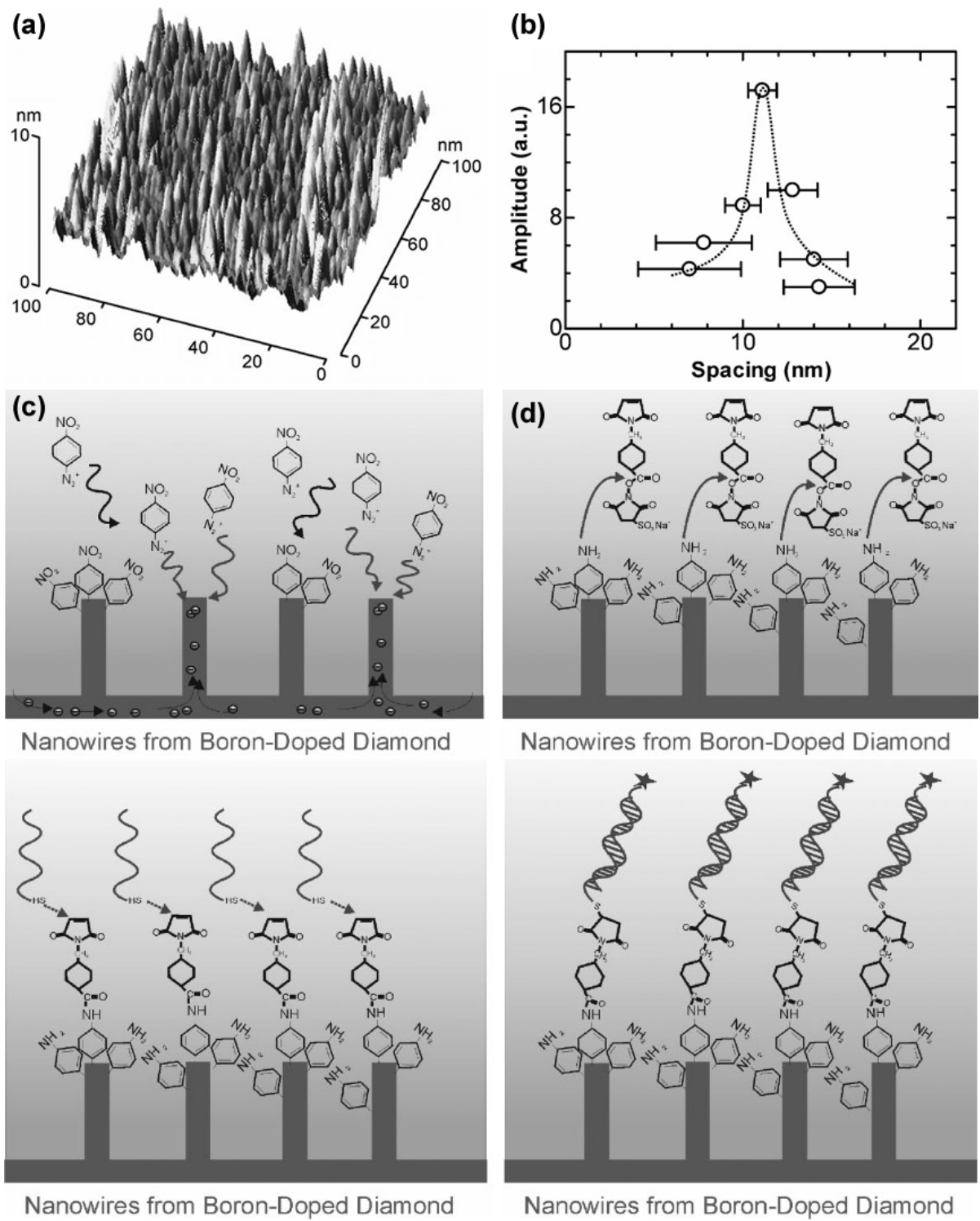

**Figure 8**. (a) Typical AFM image of diamond nanostructured surface; b) fourier transformed surface properties of diamond nanostructured surface. From Fourier analysis, the average wire separation is about 11 nm with a narrow variation; (c) – (f) Schematic pictures of the biofunctionalization of vertically aligned diamond nanowires: (c) electrochemical grafting of nitrophenyl, (d) amination and crosslinker attachment, (e) probe DNA attachment, (e) DNA hybridization.

## Energy Storage Based on Nanostructured Diamond

Due to the direct link between the electrical double layer capacitance and the surface enlargement, it is reasonable to use surface enlarged electrodes for energy storage devices. The idea is even more rationalized by the fact that the energy storage in double layer capacitors is proportional to the square of potential window: [104]

$$E = \frac{1}{2}CV^2 \qquad (5)$$

$$P = \frac{V^2}{4R_s}, \qquad (6)$$

where $E$, $P$, $C$, $V$ and $R_s$ are the energy, maximum power, capacitance, potential window and series resistance, respectively. As is pointed out, the diamond has so far the widest reported potential window reported in the aqueous electrolytes. Therefore, the attempt to use diamond as a potential supercapacitor material had already been made when the first known boron-doped diamond porous electrode was fabricated: Honda et al reported the investigation of the diamond nanohoneycomb electrode for double layer capacitor applications in 2000 [105] . Using cyclic voltammetry and impedance methods in a three-electrode setup, they estimated that the double layer capacitance is ~200 times higher than a flat BDD electrode. The gravimetric capacitance was calculated to be 16 F g$^{-1}$, which is trivial, compared to sp$^2$ carbon based materials at that time [106]. Therefore, a generally negative conclusion was made on the idea of diamond-based supercapacitors. After that, the research on this topic is almost blank for about ten years until 2009. Kondo et al made another attempt by using free standing boron-doped hollow fiber film for as supercapacitor electrode [38]. This time, a gravimetric capacitance of 13 F g$^{-1}$ was measured. This value is still small compared to sp$^2$ carbon materials [104]. Although the fabrication of the material is considerably easier than the previous results, there was no improvement in terms of gravimetric capacitance. Another prominent problem seen from this research is the large $R_s$. No information about the boron concentration in the diamond was given in the paper. However, the impedance spectroscopy shows an $R_s$ in the order of $10^4$ Ω cm$^{-2}$, which shows the poor conductivity of the material.

During the last four years (2012 – 2015), the topic of diamond-based supercapacitor gained popularity due to a large variety of porous, surface-enlarged diamond electrode has been fabricate, including diamond foam [43, 107], diamond hollow fibers [47, 48], diamond-coated CNTs [103, 108] and diamond-coated conductive polymers [49]. These electrodes share some common properties like easy fabrication, applicable to large scale fabrication (up to 6-inch), high boron concentration (confirmed by Raman spectrum or electrochemical measurements). A summary on the fabrication of properties of these materials are shown in **Table 1**.

**Table 1.** Comparison between different diamond materials in literature

| Nanostructures | Fabrication Method | Thickness* (μm) | Surface-Enlargement | Areal Capacitance (mF cm$^{-2}$)** | Ref. |
|---|---|---|---|---|---|
| Honeycomb | RIE etching with AAO mask | 0.5 | 200 | 1.97 | [105] |
| Hollow Diamond Fibers | Templated growth on quartz fiber filter | ~20 | ~20 | - | [38] |
| Diamond Foam | Templated growth on quartz spheres | 7 | ~40 | - | [43] |
| Diamond-Coated CNTs | Templated growth on CNT | 3 | 116 | 0.58 | [108] |
| Diamond Nanowires | ICP etching with Ni nanoparticle mask | 1 | ~10 | - | [6] |
| Diamond Foam | Layer-by-Layer Templated growth on quartz spheres | 2.6 | - | 0.598 | [107] |
| Diamond-Coated CNTs | Templated growth on CNT | ~40 | ~450 | - | [103] |
| Diamond-Coated Silicon Nanowires | Templated growth on Si Nanowires | ~5 | 13 | 0.105 | [109] |
| Diamond-Coated Porous Polypyrrole | Ultralow-temperature Templated growth on Si Nanowires | ~10 | ~300 | 3 | [49] |
| Hollow Diamond Fibers | Templated growth on quartz fiber filter | ~50 | - | 0.688 (based on two-electrode measurements) | [47] |

* For aligned wire-structures, heights of the structures are shown.

** Listed values are obtained in 3-electrode measurements unless otherwise noted.

In 2015, the result published by Gao et al shows the first diamond-based pouch-cell supercapacitor device based on free standing diamond paper [47]. The structure of the cell is shown in **figure 9 (a)**. Glass microfiber filters (GF/A, Whatman) was used as the separator and 3 M NaClO$_4$ was used as the aqueous electrolyte. The image of the device is shown in **figure 9 (b)**. Thanks to the two electrode device, many properties such as the potential window, series resistance, power and energy can be measured and calculated more reliably. The result shows a new understanding of the potential window of diamond. Instead of saying generally that diamond has a large potential window in aqueous solutions, the researchers find that it makes more sense to relate the potential window to the columbic efficiency of the device which is given by:

$$\text{Efficiency} = \frac{Q_{discharging}}{Q_{charging}} \times 100\% \quad , \quad (7)$$

where $Q_{charging}$ and $Q_{discharging}$ are the amount of charge during charging and discharging processes, respectively. The result of the window opening test on the diamond device is shown in **figure 9 (c)**. It is clearly seen that the water splitting starts around 1.3 V. However, the current is still small even at a high voltage of 2.5 V due to the slow kinetics of water-splitting at diamond surface. As a result, the columbic efficiency of the device still exceeds 90% at 2.0 V (**figure 9d**). On the other hand, although slow, the water splitting do consumes charges which are stored in the device. Therefore, when used in a potential window larger than 1.3 V, long-term energy storage is not possible. However, impedance spectroscopy shows that the relaxation time, which shows the highest working frequency of the given device [110], reaches 31.7 ms. This small relaxation time is believed to be because of the high ion mobility in the aqueous electrolyte and the macroporous nature of the diamond paper. Moreover, the stability of the diamond-based device is proved in 20000 galvanostatic charge/discharging cycles. The capacitance drop only ~8%. Therefore, the device is well suitable for high-voltage and high-frequency application in aqueous solutions.

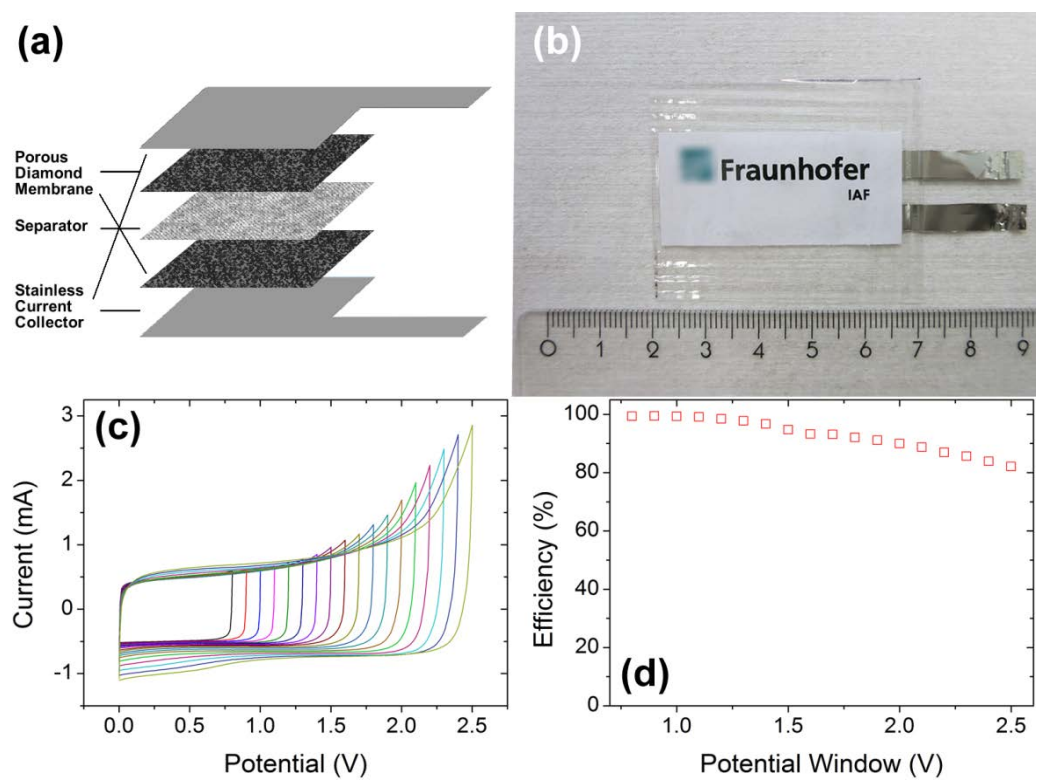

**Figure 9.** (a) Schematic illustration showing the laminated structure of a diamond-based pouch cell; (b) a photo of a diamond pouch-cell supercapacitor; the logo of the institute is intentionally blurred for copyright reasons. (c) window opening test for the diamond pouch cell in 3 M

NaClO$_4$ at 1 V s$^{-1}$ with potential windows from 0.8 – 2.5 V; (d) plot of the device efficiency against the potential window.

## Catalyst Based on Nanostructured Diamond

Nanostructured catalyst is constantly of interest for material scientist due to the nanosized effect and large specific surface [111]. However, nanosized particles are thermal dynamically unstable. Therefore, nanoparticles are often supported by other materials with also large specific surface so that the total surface energy is reduced [112, 113]. Carbon materials, often activated carbon are used for this purpose in commercially available product [114-117]. However, sp$^2$-carbon-based materials suffer from the electrochemical corrosion which is given by [118]:

$$C + 2H_2O \rightarrow CO_2 + 4H^+ + 4e^-, E^0 = 0.207 \text{ V vs SHE} \qquad (8)$$

and

$$C + H_2O \rightarrow CO + 2H^+ + 2e^-, E^0 = 0.518 \text{ V vs SHE} \qquad . \qquad (9)$$

On the other hand, diamond is resistive to electrochemical corrosion. Therefore, using diamond as a substitute for traditional porous carbon is a plausible idea.

Research on this topic has been carried out during the last decade on planar diamond electrodes [7, 119-123]. However, compared to sp$^2$-carbon-based materials, planar diamond electrode lacks the sufficient surface area which would make it a high-performance performance for catalyst. As a result, surface-enlarged diamond electrode should be used as a high-surface area support for catalyst. As early as 2001, Honda et al electrodeposited Pt nanoparticles on nanoporous diamond honeycomb electrodes [124]. In their research, electrodes with surface roughness factors of 10.9 and 15.9 were used, and Pt surfaces equals to 3 – 4 times of a planar Pt electrode were obtained. Due to the low Pt coverage on the porous electrode, the large surface area provided by the porous diamond is not fully used. Therefore, these results are comparable to later researches in which Pt nanoparticles were deposited onto planar diamond surfaces [7, 120].

In fact, the combination of porous diamond and catalytic nanoparticles has been reported only in very limited cases. The above-mentioned difficulty in uniformly depositing metal or other catalyst on the porous diamond electrode might have caused this situation. According to current results on electrodeposition on diamond electrodes, there are two main problems in achieving a uniform nanoparticle coating. One is the low nucleation density of metal deposition on diamond. Due to the inertness of a diamond surface, the nucleation can only happen on grain boundaries or other local defects [125]. In the case of a planar diamond electrode, pre-deposition treatment such as nanodiamond scratching [122], mechanical polishing [126] and wet-chemical seeding [7], have been applied to enhance the nucleation density. However, these methods are either difficult to implement on porous diamond or have not been tested so far. The second reason is the diffusion limitation in electrochemical deposition processes. During the deposition, reactive ions are depleted inside the porous matrix and the diffusing ions only arrive at the uppermost part of the porous structure. As a result, only the top of the electrode is coated, which results in a low coverage of deposits [12].

In 2015, Gao et al partially solved this problem for vertically aligned structures [127]. They applied physical instead of chemical deposition for the coating to achieve high homogeneity on high-aspect ratio structures. In their research, DC sputtering has been used to deposit thin (1 – 3 nm) Pt layers on vertically aligned diamond nanowires (**figure 10**). TEM results showed that the thin layers self-assemble into nanoparticles with diameters less than 10 nm due to the minimization of the surface energy (**figure 10c – d**). Characterization of the catalytic activities shows the diamond-Pt composite showed a high specific area of 33 $m^2\ g^{-1}$ in terms of Pt which is in the same magnitude with traditional Pt/C catalyst [117]. The electrode also achieved 23 time enlargement of Pt activity compared to a planar Pt electrode. The value is the highest reported on Pt-diamond system so far. However, for other more complex diamond structures such as diamond foam and diamond hollow fibers, to achieve a dense and uniform catalyst coating still remains as a problem to solve.

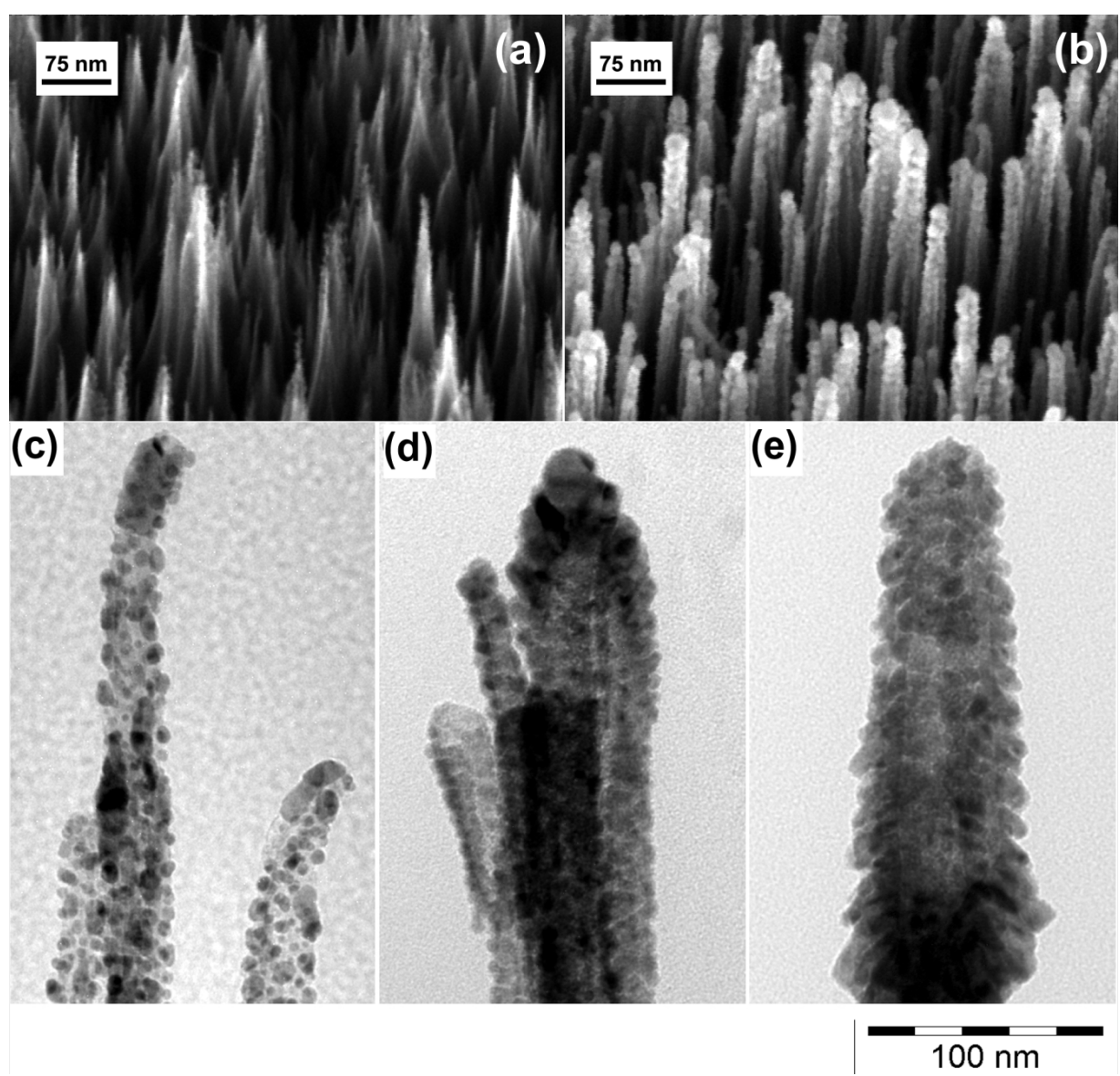

**Figure 10.** (a) SEM images of diamond nanowires and (b) diamond nanowires coated by 20 nm (nominal) Pt; (c) TEM images of diamond nanowires coated by Pt layers of nominal thicknesses of 20 nm, (d) 40 nm, and (e) 60 nm.

**Diamond Porous Membranes for Chemical/Electrochemical Separation Processes**

In recent years, due to the high demanding in robust membranes for separation and purification, researches on polymer including conductive polymers have attracted wide attention [128-130]. Compared to polymers, diamond has numerous advantages including the mechanical strength, chemical stability, wide potential window, as well as high conductivity (if highly boron-doped). The substantial development on the diamond nanostructuring during the last 15 years enables the fabrication of various porous diamond membranes with variable porosities. Therefore, the authors believe that functional diamond membranes will be an important topic in the diamond community for the coming years.

The fabrication of electrically conductive diamond membranes dates back to around the year 2000 when the through-hole diamond honeycomb electrode was fabricated [21]. However, the application was not clear until 2011 when Honda et al reported the study on an electrically-switchable diamond-like carbon (DLC) membrane [131]. The membrane was fabricated via templated growth on a porous AAO substrate. The pore size could be tuned between 14 and 105 nm. By applying a potential on the membrane, the ion flux through the membrane can be selectively accelerated or hindered depending on the charge of the ions: ions with the same charge as the membrane will be repulsed and vice versa. Although this research was carried out on DLC, the application can be easily transferred to porous diamond membranes. Due to the better conductivity, chemical stability and wider potential window of diamond, the results may be further improved.

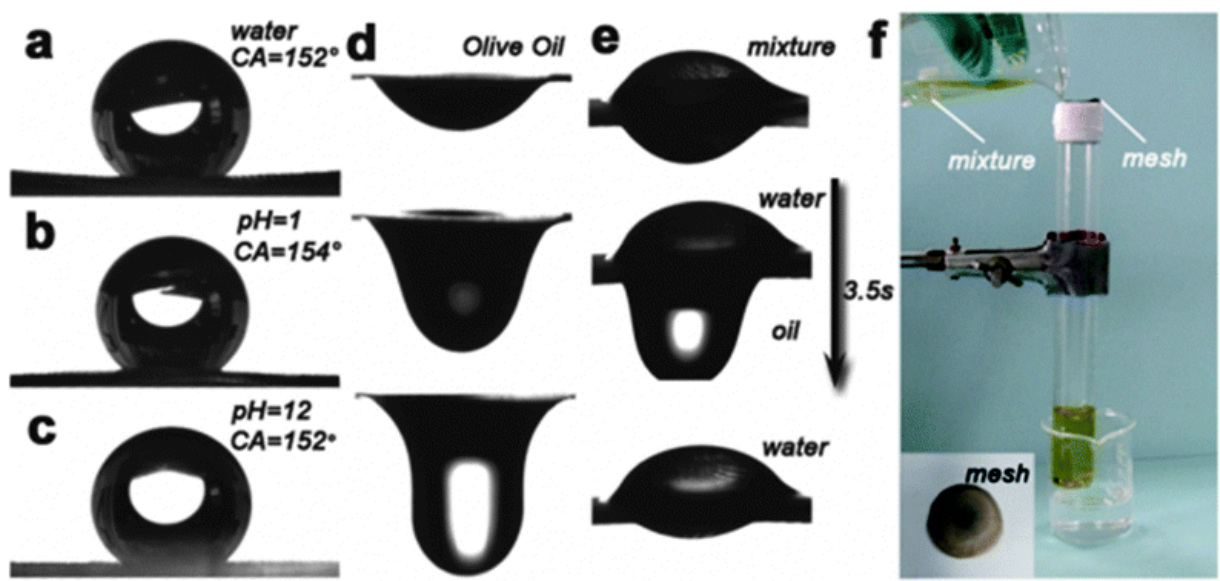

**Figure 11.** Photograph of water droplets with pH = 7 (a), acidic, pH = 1 (b) and basic, pH = 12 (c) dropped on diamond meshes showing a superhydrophobic wettability. The dynamic behaviors of a sole oil (d) and mixed water–oil droplet (e) on diamond mesh showing a quickly permeation of oil through the mesh and the separation of water–oil, respectively. (f) Photograph of a water–oil mixture separation device designed by the diamond mesh, indicating the fully water–oil separation. Reprinted with copyright permission from Cemical Communications 50 (2014), 2900—2903.

Besides potential, the surface termination of the membrane is another parameter to adjust in a diamond membrane. It has been shown that a diamond membrane can be tuned between superhydrophobic to hydrophilic by changing the surface from H- to O-termination. In 2014, Yang et al reported the fabrication of diamond membrane by coating CVD diamond on a microstructured copper mesh [132]. Due to the hydrogen rich deposition gas mixture, as-grown membranes are H-terminated and thus superhydrophobic with a contact angle of >150° for water and other aqueous solutions. By putting a droplet of water-oil mixture onto the membrane, the water will be retained by the membrane while oil will go through it. In this way, the water is separated from oil (**figure 11**). The surface-wettability can easily turn hydrophilic by air-annealing at 500 °C as the surface become O-terminated.

More functionality can be realized by further surface-modification via organic linker molecules. The surface grafting of diamond can be realized by photochemistry [87], electrochemistry [133] and wet-chemistry methods [134]. With appropriate linker molecules, it is possible to terminated diamond surface by e.g. $-NH_2$, $-COOH$, and $-HSO_3$, and different surface functionality can be realized. Ruffinatto et al showed an exemplary work on the functionalization of diamond fiber paper with aliphatic $C_4$ linkers [48]. The functionalization was realized by butylamine in an aqueous solution with a pH of 10. The mechanism is a nucleophilic substitution where the ammonium moiety acts as the leaving group. The $C_4$ functionalized diamond membrane was used in protein extraction applications. The filtering experiments showed that the functionalized diamond membrane retains 20 times more protein than the one which is not functionalized.

**Summary and Outlook**

In this chapter, the fabrication and application of a large variety of porous diamond and diamond nanostructures are introduced. Diamond fabrication techniques, including vertical structures via RIE etching and more complex 3D structures using templated-growth, has been developed in great depth during the last two decades. However, there is still room for improvements. Diamond etching method starts with thick bulk diamond. Therefore, the phase purity (in terms of non-diamond carbon) is easier to control. However, the morphology is limited to vertical structures and to achieve high aspect ratio (>30) is not yet reported. For templated-growth methods, the coatings are normally NCD with a large proportion of grain boundaries. The quality of diamond is difficult to guarantee. One solution is to achieve high density seeding so that the coalescence of nuclei happens earlier during the growth. Hopefully in this way, thinner films with lower $sp^2$-rich grain boundary proportions can be achieved. However, the $sp^2$ carbon growth in the deeper part cannot be easily inhibited. Direct growth of pure, large aspect ratio diamond wires seems to be a good solution. However, it is not yet clear if it is possible to dope these wires. Also, current reports on direct diamond nanowire growth are mainly on the report of the phenomenon; the technology is not readily on the application level.

On the other hand, the application of micro- and nanostructured diamond covers almost every aspect of electrochemistry including sensor, energy storage/conversion and separation/purification. However, challenges and chances remain on this topic. The authors would like to particularly emphasize on the diamond membrane fabrication and modification. Nowadays, the fabrication of diamond membrane has already been reported by several groups in the community with methods which are easy to handle and reproduce. By appropriate surface termination, it is expected that diamond membranes can be qualified for the tasks of other polymer membranes such as selective ion permeability and specific adsorption of chemicals. Due to the numerous unique chemical and physical properties, diamond will be a promising material for a new generation of membrane systems. Researches in this direction will lead to a broad range of new applications. On the contrary, the research on sensors based on diamond nanomaterials is less reported in recent years. As is pointed out in this chapter, the diffusion limitation in electrochemistry is a principal problem for surface enlarged electrode for sensing application. In addition, the diamond surface is inert, which means specific adsorption/accumulation of analytes is rare. Therefore, the surface enlargement cannot provide positive influence in the sensing. However, it is still possible that porous diamond can be used as porous substrate for gas sensing where diffusion limitation is not an issue.